# Binar Shuffle Algorithm: Shuffling Bit by Bit

by

William F. Gilreath
August 2008
(will@williamgilreath.com)

# Binar Shuffle Algorithm: Shuffling Bit by Bit

## Abstract

Frequently, randomly organized data is needed to avoid an anomalous operation of other algorithms and computational processes. An analogy is that a deck of cards is ordered within the pack, but before a game of poker or solitaire the deck is shuffled to create a random permutation. Shuffling is used to assure that an aggregate of data elements for a sequence S is randomly arranged, but avoids an ordered or partially ordered permutation.

Shuffling is the process of arranging data elements into a random permutation. The sequence S as an aggregation of *N* data elements, there are N! possible permutations. For the large number of possible permutations, two of the possible permutations are for a sorted or ordered placement of data elements--both an ascending and descending sorted permutation. Shuffling must avoid inadvertently creating either an ascending or descending permutation.

Shuffling is frequently coupled to another algorithmic function -- pseudo-random number generation. The efficiency and quality of the shuffle is directly dependent upon the random number generation algorithm utilized. A more effective and efficient method of shuffling is to use parameterization to configure the shuffle, and to shuffle into sub-arrays by utilizing the encoding of the data elements. The binar shuffle algorithm uses the encoding of the data elements and parameterization to avoid any direct coupling to a random number generation algorithm, but still remain a linear *O(N)* shuffle algorithm.

Keywords: permutation, randomize, shuffle, sort, unsort

## Introduction

### 1. Concept of Shuffling

Shuffling is a process of re-ordering data elements of a sequence from an initial permutation into a random arrangement of an arbitrary permutation. A shuffle algorithm scrambles data elements into a random placement without any apparent organizing key evident.

Shuffling is not as prominent or visible in the computer science domain of algorithms. However, shuffling is the inverse of sorting, a shuffle is an unsort function, and the logical complement of sorting algorithms as an unsorting algorithm.

A shuffle algorithm is conceptually similar to an unsort algorithm, thus conceptually the logical converse of a sorting algorithm. Sorting places data elements of a sequence into a very specific permutation, whereas shuffling puts the data elements into all but an ordered permutation of the possible permutations.

The correspondence between sorting and shuffling leads to a possible formalization for the definition of shuffling based upon sorting. The consideration that a shuffle algorithm is unsort algorithm, the sorting algorithm definition can be used to formulate a definition for a shuffle algorithm.

### 2. Theoretical Definition of Shuffling

Shuffling is unworkable to define in the form of an unsort algorithm--the converse definition of sorting with the formal definition of sorting using mathematical relations. However, a more tangible definition for shuffling is a probabilistic definition.

The probabilistic definition of shuffling is: Given a sequence *S* of *N* records $R_0, R_1, \ldots, R_{n-2}, R_{n-1}$ that are arranged in a permutation: p(0)p(1) ... p(n-2)p(n-1).

The sequence S is considered shuffled if for the k possible selection of any two records $R_i$ and $R_j$ where $i \neq j$, that the probability of $R_i \leq R_j$ is equal to $R_i > R_j$ for $0 < k \leq N!$ .

For the $k^{th}$ possible selection, the probability $P_k$ is:

$$P_k(R_i \leq R_j) = P_k(R_i > R_j) \text{ where } i \neq j \text{ and } 0 < k \leq N!$$

Thus after one or many selections, that any two unique records that the probability is equally likely for lesser or greater relation. No matter how many times two distinct records are selected, the overall probability of lesser or greater remains equal—there is no bias.

In contrast, a sorted sequence is defined by the probability as:

For the $k^{th}$ possible selection, the probability $P_k$ is:

$P_k(R_i \leq R_j) = 1.0$ where $i < j$ and $0 < k \leq N!$

For any two different records in a sorted sequence where the records are in increasing positions it is always true that the records maintain the lesser than relation.
It is possible to define shuffling in finite probabilities for each relation, or:

$P_k(R_i \leq R_j) = 0.5 \wedge P_k(R_i > R_j) = 0.5$ where $i \neq j$ and $0 < k \leq N!$

It is equally likely for lesser or greater for any $k^{th}$ number of selections from the sequence.

Shuffling is a random permutation of the records in a sequence, thus for the selection of records the two relations are equally likely for one, two, three, or many selections.

Conversely, sorting is an ordered permutation so for two records within the sequence at increasing or decreasing positions, the relation of lesser and greater than, respectively is always probabilistically certain.

For a sorted sequence, the probabilistic definition of shuffling is invalid, and vice-versa. Hence the probabilistic definition of shuffling is converse with sorting.

3. Approaches to Shuffling

One approach to shuffling, is to utilize the sequence generated by one pseudo-random number generator is used in the shuffle for a sequence of data elements. One proposed approach [MacLaren and Marsaglia 1965] is using the sequences of two pseudo-random number generators to improve the random number generator properties. In effect, to shuffle to more fully randomize the stochastic property of a sequence generated by a random number generator. But such an approach need not be restricted to improving the quality of a random number generator. Using randomly generated values for a sequence that is then sorted is an obvious and intuitive approach to shuffling.

An improvement to the approach of MacLaren and Marsaglia using only a single random sequence was devised [Bays and Durham 1976] and is known as a Bays-Durham shuffle, or sometimes Bay's shuffle. While an improvement, it still couples the random number generation and the specific algorithm used into the shuffle. In the two algorithmic approaches, the shuffle was to improve an already existent random number generator, hence the random number generation was already a part of the algorithm. With a shuffle of data elements to randomize or scramble the elements into a stochastic permutation, coupling to the performance complexity of the random number generator is a flawed approach.

It seems almost inevitable that a random number generator algorithm is coupled, and becomes part of the composition of shuffling. The shuffle involves organizing the data elements, but this algorithm is then used as a composite with the random number generation algorithm to create the shuffle algorithm.

Knuth gives a shuffling algorithm Algorithm M [Knuth 1998] In Algorithm M, there is a step:

M2. [Extract j.] Set j ≤kY/Mƒ, where m is the modulus used in the sequence ⟨$Y_n$⟩ ; that is, j is a random value 0 ≤ j < k, determined by Y.

This step in the algorithm couples the shuffle algorithm to the random number generator. The shuffle algorithm formulated by Bays and Durham [Bays and Durham 1976] is an optimization of Algorithm M, but has the same algorithmic step as Algorithm M.

Later, Knuth describes the Fischer-Yates algorithm what he calls Algorithm P. There is a step:

P2. [Generate U.] Generate a random number U, uniformly distributed between zero and one.

Once again, the shuffle algorithm is coupled to the random number generator algorithm.

An improvement, and an implicit constraint is that a shuffle algorithm is independent of the permutation of data elements. More simply, the shuffle algorithm is not directly coupled to a particular random number generation algorithm.

4. Summary

Shuffling is like an unsort algorithm conceptually; however, it is difficult to define shuffling strictly as the logical inverse of a sorting algorithm. A more workable approach to formalizing the concept of shuffling is to use a probabilistic definition. The distinction in formalization highlights the primary different in shuffling to sorting. Sorting has a mathematical relation among each data element in the sequence, but shuffling has a probabilistic relation among all the data elements.

Algorithm Synopsis

The binar shuffle algorithm utilizes the encoding of the data elements, more specifically the bits, to partition the data elements into a random organization. A data element is encoded with some arrangement of the bits in particular structure. For each bit, the binar shuffle places the element in a lower sub-array for a 0-bit, or the element in an upper sub-array for a 1-bit—partitioning the array of data elements.

Consider an ordinal byte, an encoding of eight bits from a most significant bit (MSB) to a least significant bit (LSB). Following the most significant bit to the least significant bit, and moving the data elements into lower and upper sub-arrays, respectively would create an ordered arrangement of the data elements—the binar sort. The sequence of bits for the encoding is the encoding order. However, to shuffle, the bits are used to place the data elements in a random arrangement by randomly accessing the bits from 0 to 7 of the 8-bits--but not in the encoding order, or the reverse encoding order. The 8-bits are put into a bit schedule for how each bit it used to place the element into one of the sub-arrays.

The M-bits of a data element are scheduled by the index position of the bits and the bit value itself. This bit schedule is used to shuffle the data elements, for a given bit within the data element, and a bit value. The bit schedule is passed to the binar shuffle as two arrays, one for the specific bit value at an index, and the other for the bit position at an index.

For an array of N-data elements, partitioning of an array into two sub-arrays continues for each bit in the encoding in a data element. Thus for an 8-bit array of byte data elements, there are $8 \cdot N$ passes for a given bit schedule to shuffle the array.

## Operation of Algorithm

The binary shuffle is exactly the same in the process as the binar sort, so that algorithm operation discussion is used. The difference in the binar shuffle to the binar sort is that each bit is used to re-arrange into a non-ordered placement of data elements. Each bit is not used as a key to map a data element into a sub-array, but is compared against the bit schedule for placement to form a random permutation.

The binar shuffle algorithm operates both iteratively and recursively in place on the original array passed as a parameter initially. The algorithm operates in four discrete steps similar to the binar sort, which are:

1. Evaluate for recursive base case.
2. Initialize starting array bounds.
3. Partition array into one or two sub-arrays.
4. Determine recursive call on sub-arrays.

## Evaluate for Recursive Base Case

The first step is to evaluate the passed parameters for termination of the algorithm, to determine if a recursive base case has been reached. When one of the two possible base cases of the binar shuffle is reached, the recursion terminates, and the call returns without any further operation on the passed parameters. The two criteria for the base case of the recursion are:

1. Reach the end of the bits in an element.
2. The size of a sub-array is one element.

Both cases are simple enough. The first case is to reach the end of the number of bits for a given data element. In effect, there are no more bits to extract to use for partitioning. The second case the bounds of the array parameter are evaluated to see if there are any elements to partition into a sub-array. For an array of one element there is no point to partition, as the element is in its final position. Thus for either case the operation of the binar shuffle terminates, or the recursive method call returns.

## Initialize Starting Array Bounds

If the binar shuffle algorithm does not terminate, then the operation proceeds to the initialization step. The passed array bounds are used to initialize the bounds for the original array. The passed array bounds are retained for use later in the operation of the algorithm. The bounds of the original array are used by variables to track the changing boundaries of the original array during partitioning. After initializing the original array boundaries, the operation then proceeds to partition.

## Partition Array into Sub-arrays

The heart of the binar shuffle algorithm and the key operation is the partitioning of the original array. The partition step of the algorithm divides the elements of the original array into lower and upper sub-arrays. The partition operation has three distinct steps:

1. Extract the nth bit from the data element in selected position.
2. Using the bit value, place the data element in the sub-array.
3. Adjust array boundary so the used sub-array is extended to encompass element.

It is important to note that the selected position used in the operation is the lower position of the array. The selected position acts as a point of focus for the operation of partitioning the array, the data element at the lower position is the working element, the element that is being placed into a sub-array by partitioning.

## Bit Extraction

Before any partitioning is possible on the selected data element, the bit at the particular position in the bit schedule must be extracted to determine the sub-array to place the element. The process of bit extracting uses a shift operation to the left by N-bits, and a bit mask to extract the bit as an integer zero or non-zero (not necessarily integer value of 1, but the integer value of the bit mask literal). The bit mask is a literal that depends on the data element word size. The bitwise logical and operation masks all the bits to zero or to the value of the literal bit mask.

## Placement of Element

The placement of the data element is dependent upon the bit value from the bit extraction. Depending on the bit value and the bit in the bit schedule, the data element is placed by one of two possibilities. The data element selected is in the lower sub-array. The two possibilities are:

1. The data element is already placed in position in the lower sub-array.
2. The data element is in wrong position, exchange with the element in the upper sub-array.

For a bit value that equals the schedule bit, the data element is in position in the lower sub-array, thus nothing is done. For a bit value that is unequal to the schedule bit, the data element is exchanged or swapped with the data element in the upper sub-array. In either case, the array bounds are adjusted afterwards once the data element is placed.

## Adjust Array Bounds

Once the data element is placed in the correct sub-array, the array boundaries are adjusted to encompass the element. For the lower sub-array, the lower array bound is incremented, and for the upper sub-array, the upper array bound is decremented. In the operation of the algorithm, the lower and upper array bounds approach one another as each data element is placed in the correct sub-array.

Partition Repetition

The partitioning process continues iteratively for each data element to be correctly placed in a sub-array. The iterative process continues until the array bounds cross over or overlap. At which point all the elements are partitioned into the sub-arrays. The last step is to determine the recursive method invocation to continue the operation of the algorithm on the sub-arrays.

Determine Recursive Call on Sub-arrays

The last primary step of the operation of the binar shuffle is to continue the algorithm recursively on the sub-arrays. Depending upon the process of partitioning, it is possible that one or two sub-arrays were created. For one sub-array, no partitioning occurred; effectively the original array is undivided. The condition of not partitioning the original array into two sub-arrays is termed pass-through. With two sub-arrays, the partitioning process successfully divided the data elements of the original array into two sub-arrays.

However, the operation of the algorithm must determine which case is the result of partitioning--pass-through with one sub-array, or two sub-arrays. The original passed array bounds are evaluating using the array bound variables that changed during partitioning. From the evaluation the recursive call is either a single recursive call or two recursive calls. The parameters passed involve the original array bounds, the array variables modified during partitioning, and the original array. For either recursive call the bit position is the next incremental bit position in the data element from the current position $i$ to the next position $i+1$.

Illustration of the Algorithm

The binar shuffle algorithm works on different data types (such as character, integer, ordinal, float, double, string) of different data sizes. The illustration of the binar shuffle is for example values moved into the lower or upper sub-arrays for a shuffle.

Consider an initial array of 32-bit word values [0 1 2 3 4 5 6 7 8 9 10 11 12 13 14 15] that are sixteen 32-bit words to shuffle into a random permutation. The bit mask for the bit extraction is hexadecimal 0x00000001 or integer 1. A total of 8-bits are utilized from the 32-bit integer word, from bit positions 31 to 24. The bit values are an alternating pattern of one and zero or the series 1,0,1,0,1,0,1,0 respectively. The initial array using the binar shuffle algorithm is partitioned into a new random permutation of [12 11 6 7 8 9 10 0 15 14 13 2 3 4 5 1] for the final shuffled array.

The initial call is of the form:

```
shuffle(0, data.length-1, 0, data, indx, bits, 8);
```

The call to the method shuffle, with the sub-array from 0...length of the data array, an initial index position of 0, the data sub-array (the initial array passed), the index of bit positions, the bit schedule, and the number of bits to use in shuffling the array.

There are several recursive invocations, but to illustrate the initial first pass to partition the array into sub-arrays is used for illustration. Initially the two resulting sub-arrays are empty, and the starting array contains all 16-data elements. The array and sub-arrays have the data elements at the start:

```
Array = [][0 1 2 3 4 5 6 7 8 9 10 11 12 13 14 15][]
```

The first value selected for partitioning is the data element integer value of 0. The extracted bit is a 0-bit, and the bit value for the bit schedule is a bit-0 is equal. Thus the integer value of 0 is in the correct lower sub-array. In this case, the sub-array boundaries are incremented to include the data element, as there is no data element exchange. The resulting arrays are:

```
Array = [0][1 2 3 4 5 6 7 8 9 10 11 12 13 14 15][]
```

The second value selected for partitioning is the data element integer value of 1. The extracted bit is a 1-bit, and the bit value for the bit schedule is a bit-0. The extracted bit from the scheduled bit is not equal, thus the data element is exchanged with the data element in the upper sub-array. The resulting arrays are:

```
Array = [0][15 2 3 4 5 6 7 8 9 10 11 12 13 14][1]
```

The next data element partitioned is the integer value 15, and so forth. This process of partitioning into sub-arrays of shuffled data elements using the bit value extracted continues recursively. The final permutation of the array is:

```
Array = [12 11 6 7 8 9 10 0 15 14 13 2 3 4 5 1]
```

The initial array of sorted elements partitioned into a final permutation that is randomly arranged. One important point in illustrating the operation of the binar shuffle is that there are several parameters to this particular configuration of the algorithm. The initial permutation of the data *d'*, the data element bit size *w*, the number of bits used *n*, the bit schedule *s*, and the bit indices *i*. Another unique final permutation from the binar shuffle is possible or a parameterized form of *B(n,s,i)* where the 3-parameters of number of bits used *n*, the bit schedule *s*, and the indices *i*.

Analysis of Algorithm

The binar shuffle is in performance linearly proportional in both time and space, or Big-Oh of *O(n)*.

a. Space

The space performance, or memory utilized by the binar shuffle algorithm is the simplest analysis. The binar shuffle uses the data of an indices and bit schedule, along with the number of bits to use, and the starting index position. The indices, bit schedule, and number of bits to use are read-only constants never updated. The index position is updated, as are the sub-array boundaries. The starting array of data elements, and following sub-arrays are the same array structure, passed recursively. Hence the size of data elements is *N*, and the bit schedule and indices are of size *N*. The other parameters passed, and used within the binar shuffle are a constant number of *c*.

The space performance complexity is the sum of the two quantities--the arrays and the variables. The three arrays are of size *N*, and thus in Big-Oh notation are $c{\cdot}N$. The variables are a constant number *c*; for the expression of the sum that is the space complexity is the arithmetic expression $c{\cdot}N+c$. Using Big-Oh notation, the space analysis performance is $O(c{\cdot}N+c)$ or more simply $O(N)$ that is linear space complexity.

b. Time

The time performance or runtime performance of the binar shuffle algorithm is more involved. The partitioning recursively into sub-arrays recursively has the possibility of complex mathematical analysis.

Before delving deeper into the analysis of the time performance of the binar shuffle, there are several important specific points to consider. The three points that relate to time performance are:

1. Each bit at an index is used only once to shuffle.
2. Each data element for a bit is accessed only once.
3. For a given data element bit size of *w*, only *s* bits are used where $0 < s \leq w$.

For an array of N-elements, there are a constant *c* number of bits to use, the given size *s* of bits. Each bit extracted is used once in the shuffle, and a data element has its bit accessed once. Thus the time performance is a product expression of the number of bit accesses *a*, the number of bits used *c*, and the number of data elements *N*. The time complexity is expressed in the form of an arithmetic expression of:

$$T = a{\cdot}c{\cdot}N$$

Since each bit is accessed once for each pass to shuffle the data elements, the time performance expression is simplified to:

$$T = 1 \cdot c \cdot N = c \cdot N$$

Thus the time performance is linear or $O(c \cdot N)$ or more simply $O(N)$.

A different examination and analysis of the time performance is that the array of N-elements and using s-bits to shuffle is a matrix of N-rows and s-columns. The size of $N$ can vary, but the number of bits $s$ is a constant $c$. The algorithm accesses the $N \times c$ cells of the matrix for each row and each column once. Thus to access all the cells in the matrix once will have a time performance complexity of $O(N \cdot c) = O(c \cdot N) = O(N)$.

Performance of Algorithm

A. Test of Performance

The binar shuffle algorithm was implemented in the C programming language, and compiled with the GNU C Compiler (gcc) for the PowerPC platform. The test of performance of the binar shuffle algorithm varies in size from the initial size to ten times the original size. A shuffle algorithm randomizes data, thus the permutation of the elements of the test set need only be in sorted or ordered organization. The test data set uses unique, non-repeated data elements to avoid any potential anomalous data elements, and to test the performance of shuffling on each data element in a data set only once.

The test program creates a data set that consisted of ordered 32-bit integers, varying from an initial size of 200,000-elements to 2,000,000-elements in increments of 200,000. The test data set was ordered in both ascending and descending order, and the integers were non-repeated. The test program was executed several times to avoid any spurious inconsistencies.

B. Performance Results

The results of the performance tests on the generated data sets were consistent with theoretical analysis.

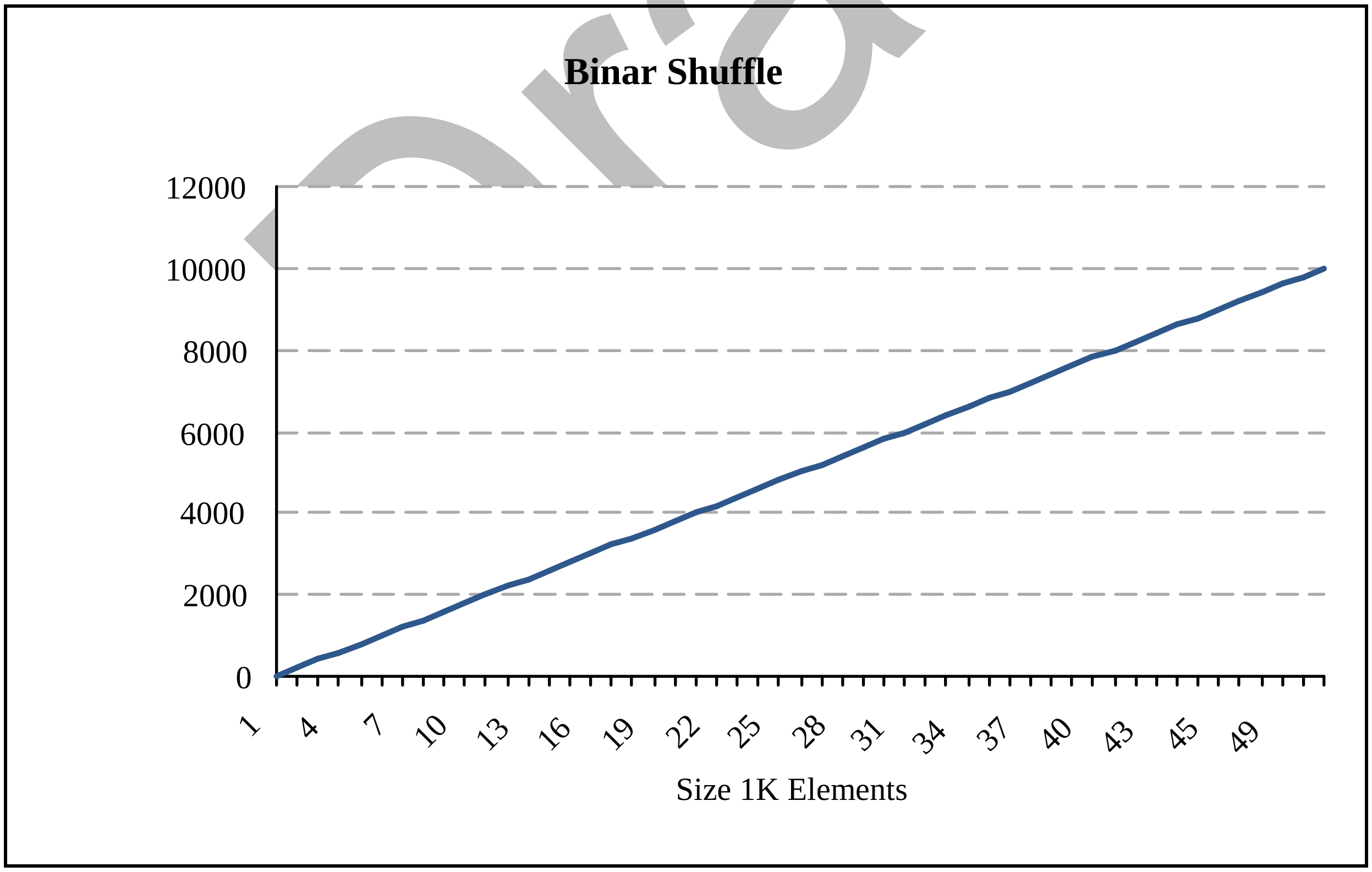

Graph of Size to Performance of Binar Shuffle

One unexpected result was that as the data set size increased linearly, the performance time increased linearly in proportion, but at a sub-linear rate. However, this sub-linear time to data size is still the Big-Oh complexity of *O(N)* , but with a constant $c < 1$. The constant varied depending on the size of the test data set, but remained within the constraint of $0.417 \leq c \leq 0.5$, or more specifically $0 < c < 1$.

The binar shuffle algorithm performance is consistent with the theoretical analysis as a Big-Oh *O(N)* linear algorithm. The constant flux depends on the test data set size but is a consistent constant for linear performance.

## Future Work

The binar shuffle algorithm is by no means finished or done, there are much further work and possibilities with the algorithm. Future work with the binar shuffle has the goal of improving the performance of the binar shuffle, optimizing the implementation of the binar shuffle algorithm, and potentially generalizing the binar shuffle for many data types and encodings.

All three venues for future work must avoid creating a non-linear algorithm, thus preserving performance in both time and space. All the while for the optimizations, the binar shuffle must avoid the anathema of a shuffle algorithm--creating an ordered or partially ordered permutation, inadvertently sorting the array of data elements. These constraints form the bounds for future work with the binar shuffle algorithm.

There are several potential areas for future work with the binar shuffle algorithm that are:

1. Variations of the algorithm
2. Bit scheduling
3. Parallelization
4. Dynamically adjust for re-shuffle

These four areas named for further endeavor are not the only potential prospects, but are apparent avenues for more future work.

### A. Variations

Variations of the binar shuffle algorithm are mainly in two variants of the binar shuffle algorithm. These two changes are:

1. Translation of binar shuffle to other programming languages and paradigms.
2. Iterative implementation of the binar shuffle algorithm.

#### a. Translation to Other Programming Languages

One area of future work is to port the binar shuffle to other paradigms of programming languages. Often an algorithm is easily implemented in one programming language and paradigm, but it is much more difficult to translate and port the algorithm to another programming language and paradigm. However, (famous last words before eating them) it is not impossible, as every programming language is equivalent in Turing completeness. Hence in theory the binar shuffle algorithm can be translated and ported to any other programming language, but in practice it might be more or less difficult to do so in actuality.

One paradigm of programming of interest is that of functional programming languages. The interesting point of consideration is utilizing the functional paradigm to implement the binar shuffle, and how the binar shuffle algorithm operates using a functional approach to its implementation. It would be equally interesting to port the binar shuffle to other programming languages that take a different approach to implementation, and have unique and novel features.

The question is one of how does the binar shuffle algorithm change in comparison to other implementations is answered by translation. The answer requires actual translation to evaluate the new version of the binar shuffle, but for different implementations in different paradigms, much better insights into the binar shuffle can be determined.

b. Iterative implementation of the algorithm

The binar shuffle algorithm is primarily a recursive algorithm, after each shuffle on a bit at an index; the created sub-arrays are then recursively shuffled for the next bit index at the successive position. However, recursion can be costly, as recursion creates stack frames on the runtime stack.

One potential improvement is to use an iterative rather than a recursive approach, begging the question of if it is possible to convert the recursive binar shuffle to an iterative binar shuffle. The question remains that if it is possible, how does it alter the binar shuffle algorithm implementation.

Another optimization is to use iteration partially, if a full conversion to complete iteration is not viable. In the case that a shuffle only creates a single sub-array, rather than continue recursively, the next shuffle for a bit at an index is handled iteratively. This is a partial optimization to avoid the overhead of recursion when there are not two sub-arrays to partition with a shuffle.

B. Bit Scheduling

Future work with bit scheduling involves two potential improvements for the operation of binar shuffle that uses the bits to shuffle the data elements. The two possible improvements are:

1. Reverse bit schedule from the encoding
2. Ubiquitous bit schedule for any encoding

Reverse Bit Schedule

For a given data set of elements, the elements have a particular encoding. That is to say that the encoding follows a specific bit index, an order of the bits, within a data element from the most significant bit (MSB) to the least significant bit (LSB). When the bit schedule follows the bit order, effectively an ordered permutation is generated--the same as the binar sort.

Thus following the reverse bit schedule, or from the least significant bit to the most significant for a given encoding is a potential generalization of the binar shuffle. This optimization would eliminate the need to pass a bit index as part of the bit schedule for a given data element type. The bit index would be coded as part of the binar shuffle algorithm for the specific type of data element passed for randomization. The trade-off for this improvement is that a specific type in a specific encoding would require a binar shuffle implementation for that type and encoding.

b. Ubiquitous Bit Schedule

Improving and optimizing the bit schedule for a shuffle improves the efficacy of the binar shuffle algorithm by ensuring that for any possible permutation that the algorithm will never by design create an ordered or partially ordered permutation of the data elements.

C. Parallelization

Parallelization of the binar shuffle involves creating a parallel variant of the binar shuffle. The two possible (and again, by no means the only viable) approaches to parallelization are:

1. Use the created of sub-arrays.
2. Utilize each bit at an index.

a. Use the created sub-arrays

Parallelize the operation of the binar shuffle; at some point a sub-array, and a single processor handles the other created sub-arrays. The sub-arrays are mutually exclusive of one another, thus can shuffle recursively independent of other sub-arrays. Once all the processors finish shuffling the sub-arrays recursively, each is gathered into the original array.

b. Utilize each bit at an index

Each bit for a shuffle is mapped to a specific processor in the parallel system. Thus parallelize the binar shuffle for each of the bits at an index used.  Each shuffle uses a specific bit that is independent from the other bits in the data elements. At some point when each processor finishes shuffling using the bits, the shuffled arrays are gathered into a complete array.

The gathering of the shuffled arrays on each processor is more involved, as each bit index creates a unique permutation on each processor. The last serial process of the parallel binar shuffle is integration of each data element from each permutation on each processor into a final permutation of the array.

Both approaches at parallelization have some serial algorithmic process in the binar shuffle algorithm. An open research question is for future work involving parallelization that minimizes or avoids a serial approach in the parallelization of the binar shuffle algorithm.

D. Dynamic Adjusting for Re-shuffle

One possible area of further work is dynamic parameterization; the binar shuffle algorithm dynamically re-adjusts the parameters of the shuffle. Hence for a given final permutation, the binar shuffle will re-shuffle should the shuffle not be optimal for the parameters originally passed. Such a self-adjusting binar shuffle would re-configure the parameters of the number of bits used, and the bit schedule of indices and values. The open question is what criteria are used to determine if the final permutation needs to be re-shuffled under different parameters.

Two possible metrics for re-shuffling the array are:

1. Compare the final permutation of data elements compared with original positions in the array.

2. Analyze the final permutation to determine if any triple of elements is ordered, or sorted, such as $a \le b \le c$.

The shuffle is either repeated on the original array of data elements, or on the final permutation for the array of data elements. An automatic re-shuffling would avoid any partially ordered, incompletely shuffled, or ordered permutations of the data elements.

Other future work for improvements and optimizations is possible in the context of applications and libraries that utilize the binar shuffle.

## Conclusion

The binar shuffle is an *O(N)* linear, universal, recursive shuffling algorithm. The binar shuffle utilizes the universal property of the encoding of the data elements for shuffling. The bits for encoding of the data elements are used on the array from an initial starting permutation into a randomly arranged permutation. The extraction, exchange, and partitioning of the array into sub-arrays is the shuffle process. The binar shuffle is universal, utilizing the encoding of data elements instead of other a prior property of the data.

The binar shuffle algorithm is a configurable or parameterized shuffle algorithm. The specific operation of the algorithm is configured by a 3-tuple of the number of bits from the encoding to use, the bit index, and the bit values. The bit index and bit values form the bit schedule used for the shuffle.

The use of parameterization separates the binar shuffle from bias to a pseudo-random number generator. Instead, the binar shuffle algorithm uses a bit schedule--an index of bit positions, and bit values to organize an array of data elements into a random permutation. The random aspect of the shuffle is external to the algorithmic operation.

## Appendix A - Java Source Code

```
public final class BinarShuffleTest
{
    private BinarShuffleTest(){}

    public final static void shuffle(final int lo_bound, final int hi_bound,
                                     final int pos,      final int[] array,
                                     final int index[],  final int[] bits,
                                     final int size)
    {

         System.out.print(">>> lo = "+lo_bound+"; hi = "+hi_bound);
         System.out.print("Index: "+pos+"; Position: "+index[pos]);
         System.out.println(" <<<");
         System.out.println();

         if(pos > size || lo_bound >= hi_bound) return;

         int lo = lo_bound;
         int hi = hi_bound;

         while(lo < hi+1)
         {
             final int bit = (index[pos] >> array[lo]-1) & 0x00000001;

             System.out.print(">>> lo = "+lo_bound+"; hi = "+hi_bound);
             System.out.print("Index: "+pos+"; Position: "+index[pos]);
             System.out.println(" <<<");
             System.out.println();
             printArray(array);

             if(bit == bits[pos])
             {
                 lo++;
             }
             else
             {
                 int temp  = array[hi];
                 array[hi] = array[lo];
                 array[lo] = temp;
                 hi--;

             }//end if

         }//end while

         if(lo == hi_bound + 1)
         {
             shuffle(lo_bound,hi_bound,pos+1, array, index,bits,size);
         }
         else
         {
             shuffle(lo_bound, lo-1, pos+1, array,index,bits,size);
             shuffle(lo, hi_bound, pos+1, array,index,bits,size);
         }//end if

    }//end shuffle
```

```
    public final static void printArray(final int[] array)
    {
        System.out.print("Array = [");
        for(int x=0;x<array.length;x++)
        {
            System.out.print(" "+array[x]);
        }//end for
        System.out.println(" ]");

    }//end printArray

    public final static void main(String[] args)
    {


        int[] data = new int[]{  0,  1,  2,  3,
                                 4,  5,  6,  7,
                                 8,  9, 10, 11,
                                12, 13, 14, 15  };

        int[] indx = new int[]{ 31, 30, 29, 28, 27, 26, 25, 24,
                                23, 22, 21, 20, 19, 18, 17, 16,
                                15, 14, 13, 12, 11, 10,  9,  8,
                                 7,  6,  5,  4,  3,  2,  1,  0 };

        int[] bits = new int[]{  0, 1, 0, 1, 0, 1, 0, 1,
                                 0, 1, 0, 1, 0, 1, 0, 1,
                                 0, 1, 0, 1, 0, 1, 0, 1,
                                 0, 1, 0, 1, 0, 1, 0, 1 };

        System.out.println();
        printArray(data);
        System.out.println();

        shuffle(0, data.length-1, 0, data, indx, bits, 8);

        System.out.println();
        printArray(data);
        System.out.println();

        System.exit(0);

    }//end main

}//end BinarShuffleTest
```

## Appendix B - C Source Code

```
#include <stdio.h>
#include <stdlib.h>

#define WIN32 1
#define TRACE 0

#ifdef WIN32
     #include <time.h>
#else
     #include <sys/time.h>
#endif

#ifdef WIN32
    clock_t start, stop;
#else
    timeval start, stop;
#endif

#ifdef WIN32
        double duration(const clock_t first, const clock_t last)
        {
                return(double)(last - first) / CLOCKS_PER_SEC;
        }
#else
        double duration(timeval first, timeval last)
        {
            return (double) (1000000* (last.tv_sec  - first.tv_sec )
                                    + (last.tv_usec - first.tv_usec));
        }
#endif

void printArray(const int array[], const int len)
{
    int x = 0;
    printf("Array = [");
    for(x=0;x<len;x++)
    {
        printf(" %d ",array[x]);
    }//end for
    printf(" ] ");
    printf("\n\r");

}//end printArray
```

```
void binar_shuffle(const int lo_bound, const int hi_bound, const int pos,
                   int array[],  const int index[],  const int bits[],
                   const int size)
{
     int lo = -1;
     int hi = -1;

     if(TRACE) printf(">>> lo = %d; hi = %d; Index: %d; Position: %d <<< \n\r",
                       lo_bound,hi_bound,pos,index[pos]);

     if(pos > size || lo_bound >= hi_bound) return; //uint = 32-bits+1 = 33

     lo = lo_bound;
     hi = hi_bound;

     while(lo < hi+1)
     {
         const int bit = (array[lo] << index[pos]) & 0x80000000;

         if(TRACE) printf("lo = %d; hi = %d ; Index: %d ; Position: %d; \n\r",
                           lo,hi,lo,hi,pos,index[pos]);
         if(TRACE) printArray(array,hi_bound-lo_bound+1);

         if(bit == bits[pos])
         {
             lo++;
         }
         else
         {
             int temp  = array[hi];
             array[hi] = array[lo];
             array[lo] = temp;
             hi--;
         }//end if

     }//end while

     if(lo == hi_bound + 1)
     {
         binar_shuffle(lo_bound,hi_bound,pos+1, array,index,bits,size);
     }
     else
     {
         binar_shuffle(lo_bound, lo-1, pos+1,array,index,bits,size);
         binar_shuffle(lo, hi_bound, pos+1,array,index,bits,size);
     }//end if

}//end binar_shuffle
```

```
void test_shuffle_ascend(const int size)
{
    double  time = 0.0;

    int x      = -1;
    int indx[] = { 31, 30, 29, 28, 27, 26, 25, 24, 23, 22, 21, 20, 19, 18, 17, 16,
                   15, 14, 13, 12, 11, 10,  9,  8,  7,  6,  5,  4,  3,  2,  1      };
    int bits[] = {  0, 1, 0, 1, 0, 1, 0, 1, 0, 1, 0, 1, 0, 1, 0, 1, 0, 1, 0, 1, 0,
                    1, 0, 1, 0, 1, 0, 1, 0, 1, 0 };

    int* data = malloc(sizeof(int)*size);

    for(x=0;x<size;x++)
    {
        data[x] = x;
    }//end for

    start = clock();
    binar_shuffle(0, size-1, 1, data, indx, bits, 4);
    stop  = clock();

    time = duration(start,stop);

    printf("Time to shuffle size %d is %f seconds \n\r",size,time);
    printf("%d,%f \n\r",size,time);

}//end test_shuffle_ascend

void test_shuffle_dscend(const int size)
{

    double  time = 0.0;

    int x      = -1;
    int indx[] = { 31, 30, 29, 28, 27, 26, 25, 24, 23, 22, 21, 20, 19, 18, 17, 16,
                   15, 14, 13, 12, 11, 10,  9,  8,  7,  6,  5,  4,  3,  2,  1      };
    int bits[] = {  0, 1, 0, 1, 0, 1, 0, 1, 0, 1, 0, 1, 0, 1, 0, 1, 0, 1, 0, 1, 0,
                    1, 0, 1, 0, 1, 0, 1, 0, 1, 0 };

    int* data = malloc(sizeof(int)*size);

    for(x=size-1;x>=0;x--)
    {
        data[x] = x;
    }//end for

    start = clock();
    binar_shuffle(0, size-1, 1, data, indx, bits, 4);
    stop  = clock();

    time = duration(start,stop);

    printf("Time to shuffle size %d is %f seconds \n\r",size,time);
    printf("%d,%f \n\r",size,time);

}//end test_shuffle_dscend
```

```
#define DELTA 200000
#define LIMIT 10

int main()
{
    int x = -1;

    printf("\n\r");

    for(x=0;x<LIMIT;x++)
    {
       test_shuffle_ascend(DELTA*x+DELTA);
    }//end for

    printf("\n\r");

    for(x=0;x<LIMIT;x++)
    {
        test_shuffle_dscend(DELTA*x+DELTA);
    }//end for

    printf("\n\r");

    return 0;

}//end main
```